\documentclass[amsmath,amssymb,aps,prl,twocolumn]{revtex4-2}
\usepackage{physics}
\usepackage{xcolor}
\usepackage{graphicx}
\usepackage{amsmath}
\usepackage{amssymb}
\usepackage{bbold}
\usepackage{bm}
\usepackage{mathtools}

\newcommand{\cor}[1]{\textcolor{black}{#1}}
\newcommand{\new}[1]{\textcolor{black}{#1}}

\begin{document}
\title{Ground state nature and nonlinear squeezing of Gottesman-Kitaev-Preskill states}

\begin{abstract}
The main bottleneck for universal quantum computation with traveling light is the preparation of Gottesman-Kitaev-Preskill states of sufficient quality. This is an extremely challenging task, experimental as well as theoretical, also because there is currently no single easily computable measure of quality for these states. We introduce such measure, \new{ GKP squeezing}, and show how it is related to the current ways of characterizing the states. The measure is easy to compute and can be easily employed in state preparation as well as verification of experimental results.
\end{abstract}

\author{Petr Marek}
\affiliation{Department of Optics, Palack\'y University, 17. listopadu 1192/12, 771 46 Olomouc, Czech Republic}
\maketitle

Universal quantum computation is a goal pursued on many experimental platforms \cite{arute2019,wang2019,zhong2020}, each one with a set of advantages and disadvantages. Bosonic harmonic oscillators encoded into modes of room temperature traveling light offer unprecedented scalability, arising from their compatibility with modern communication technologies \cite{rudolph2017,bourassa2021,inoue2023}. Traveling light has already been used for demonstrating quantum advantage \cite{wang2019,zhong2020}, but the universal computation requires passing a significant bottleneck - preparation of suitable non-Gaussian states.

Gottesman-Kitaev-Preskill (GKP) states can provide the necessary non-Gaussianity \cite{gottesman2001,baragiola2019,terhal2020,bourassa2021,grimsmo2021}. They can be effectively multiplexed and error corrected by feasible Gaussian operations and measurements \cite{miwa2014,asavanant2019,larsen2019,yoshikawa2016,konno2021,deneeve2022,inoue2023}. However, preparation of high quality GKP states is a task, which has so far been achieved only in systems with strong coupling to an auxiliary qubit \cite{fluhmann2019,campagneibarcq2020,hastrup2022}. Unfortunately, traveling light requires \new{a more challenging} approach \cite{konno2023}. The currently available optical preparation is based on the boson sampling \cite{hamilton2017}, in which  \new{some modes of a multi-mode entangled Gaussian state} are measured by photon number resolving detectors and some outcomes herald preparation of the desired state \cite{yukawa2013,weigand2018,eaton2019,eaton2022,konno2023}. A key component of this approach is the numerical simulation optimization of the state preparation setup with the goal of preparing a suitable state with a feasible success rate \cite{killoran2019,deprins2023}. However, quantifying the quality of the prepared GKP approximation is still an open problem.

The ideal GKP state is a non-normalizable abstraction with unit value of the stabilizers and infinitely squeezed states \cite{gottesman2001}. All of these are separate properties, so, rather than using a multi-parameter optimization, the current customary approach relies on devising an approximative state with limited quality and aiming to prepare it with high fidelity \cite{duivenvoorden2017,weigand2018,eaton2019,mensen2021,eaton2022,konno2023}. However, for the sake of both setup optimization and state evaluation, it would be practical to identify a single measure of non-Gaussianity inherent to GKP states, similarly to the cubic squeezing of resource states for the cubic phase gate \cite{gottesman2001,miyata2016,konno2021b,kala2022,brauer2021}. 

In this letter we introduce a class of Hermitian positive semi-definite operators, whose ground states are the GKP qubit states in various topologies. We show that their mean values can be interpreted as nonlinear squeezing of the GKP states and that this squeezing neatly ties together the existing methods for evaluation. Furthermore, it  can be advantageously used for efficiently evaluating experimentally prepared states, optimizing the state preparation procedures, and even arriving at fundamental relation between the quality of GKP states and their stellar rank.

|

The ideal GKP qubit state $|0_L\rangle$ is defined as the infinite superposition of quadrature eigenstates,
\begin{equation}\label{gkp ideal}
    |0_L\rangle \propto \sum_{s \in Z} \left|x = 2 s \sqrt{\pi}\right\rangle.
\end{equation}
Its main properties arise from the state being the simultaneous eigenstate of stabilizer displacement operators $\hat{S}_x = e^{-i2\sqrt{\pi}\hat{p}}$ and  $\hat{S}_p = e^{-i2\sqrt{\pi}\hat{x}}$,\cor{ where $\hat{x}$ and $\hat{p}$ are quadrature operators with $[\hat{x},\hat{p}] = i$.} \new{Furthermore, the state is preserved under $\sqrt{\hat{S}_p}$ and changes to the orthogonal state $|1_L\rangle$ under $\sqrt{\hat{S}_x}$. } Any approximation \cor{$|\tilde{0}_L\rangle$} of the state (\ref{gkp ideal}) should follow
\begin{align}\label{stabilizers}
\hat{S}_x|\tilde{0}_L\rangle \approx |\tilde{0}_L\rangle, ~ \sqrt{\hat{S}_p}|\tilde{0}_L\rangle \approx |\tilde{0}_L\rangle,
\end{align}
where `$\approx$' represents approximate equality and would be replaced by the equality for the ideal state $|0_L\rangle$. The main result of this paper is the realization that the GKP qubit \new{state} (\ref{gkp ideal}) can be also found as the zero eigenvalue eigenstate of operator
\begin{align}\label{Q0}
    \hat{Q}_{0} &= 2\sin^2\left(\frac{\hat{x}\sqrt{\pi}}{2}\right) + 2\sin^2(\hat{p}\sqrt{\pi}) \nonumber \\
& = \frac{1}{2}[ 4  - \sqrt{\hat{S}_p} - \sqrt{\hat{S}_p}^{\dag} - \hat{S}_x - \hat{S}_x^{\dag}].
\end{align}
\new{The assymetry between $\hat{x}$ and $\hat{p}$ follows the asymmetry of the GKP state (\ref{gkp ideal}) and the minimal operations (\ref{stabilizers}) that preserve it.}
For any quantum state, the mean value $\langle\hat{Q}_0\rangle$ directly depends on the value of stabilizers and it can be zero only when eigenvalues of both $\hat{S}_x$ and $\sqrt{\hat{S}_p}$ are equal to 1. At the same time, the asymmetry between $\hat{x}$ and $\hat{p}$ ensures that the eigenstate of $\hat{Q}_0$ is indeed the GKP qubit state $|0_L\rangle$, rather than any other state  preserved under the action of the stabilizers. Looking at it from a different perspective, operator (\ref{Q0}) can produce the value of zero only for those quantum states with non-zero wave function only for discrete periodic values in both $x$ and $p$ representation. This is a property satisfied only by the GKP qubit state $|0_L\rangle$. 

The main advantage of the specific combination of displacement operators is that the resulting operator $\hat{Q}_0$ is Hermitian and positive semi-definite. The operator can be therefore taken as a Hamiltonian and the GKP state as its ground state. Furthermore, the mean value $\langle\hat{Q}_0\rangle$ \cor{in some cases equals to the} variance of  $\sqrt{\hat{Q}_0}$, which is a nonlinear functional of quadrature operators, and can be therefore interpreted as nonlinear squeezing. The concept of squeezing is closely tied to continuous variable quantum optics and it serves as one of its main resources \cite{braunstein2005,yadin2018,carrara2020}. As a figure of merit for preparation of squeezed states, the variance of $\hat{x}$ neatly transitions between the classical limit given by the vacuum state with variance $\langle \mathrm{vac}|\hat{x}^2|\mathrm{vac}\rangle = 1/2$, and the ultimately unachievable limit of the unphysical quadrature eigenstate $\langle x =0|\hat{x}^2|x=0\rangle \rightarrow 0$. The degree of squeezing for any quantum state $\hat{\rho}$ with $\langle \hat{x}\rangle = 0$ can be then given by the normalized second moment $2\mathrm{Tr}[\hat{\rho}\hat{x}^2]$, \cor{ which can be expressed either as a direct value or in dB}. The concept can be expanded to moments of other operators, which has been demonstrated on resource states for the cubic phase gate \cite{miyata2016,konno2021,kala2022,brauer2021,provaznik2022}.

In \cor{ analogy to quadrature squeezing}, we can define the \cor{nonlinear} \emph{GKP squeezing} \new{for any quantum state $\rho$} as
\begin{align}\label{xi gkp}
    \xi_0 = 
    \langle\hat{Q}_0\rangle =  \mathrm{Tr}[\hat{\rho} \hat{Q}_0].
\end{align}
{It can be quickly shown that, for any Gaussian state, the GKP squeezing cannot be lower than one, which is achieved for a quadrature eigenstate \cor{\cite{supplement}}. Quantity (\ref{xi gkp}) is also bounded for classical states that can be expressed as mixtures of coherent states and always result in $\xi_0 \ge 2- e^{-\pi/2}-e^{-\pi}$.}
Note that definition (\ref{xi gkp}) does employ the second moment \new{of $\sqrt{\hat{Q}_0}$}  and not the variance, which is customarily considered in the Gaussian squeezing scenario. The omission of the square of the first moment was done deliberately to keep the \cor{GKP} squeezing as a linear functional of \new{the} density matrix. The choice is also justified as the optimal states have the mean values equal to zero.

The GKP squeezing (\ref{xi gkp}) is defined for the GKP state \cor{ $|\tilde{0}\rangle$}. However, since other GKP states and other encodings are only a Gaussian operation away, we can easily define operators for the other GKP states by replacing the arguments of the functions. We can thus define
\begin{align}\label{Qother}
    \hat{Q}_{1} &= 2\cos^2\left(\frac{\hat{x}\sqrt{\pi}}{2}\right) + 2\sin^2\left(\hat{p}\sqrt{\pi}\right),\nonumber \\
    \hat{Q}_{s0}& =  2\sin^2\left(\hat{x}\sqrt{\frac{\pi}{2}}\right) + 2\sin^2\left(\hat{p}\sqrt{\frac{\pi}{2}}\right), \nonumber \\
    \hat{Q}_{s1} &=  2\cos^2\left(\hat{x}\sqrt{\frac{\pi}{2}}\right) + 2\sin^2\left(\hat{p}\sqrt{\frac{\pi}{2}}\right), \nonumber \\
    \hat{Q}_h &= 2\sin^2\left(\kappa_+\hat{x}-\kappa_-\hat{p}\right) + 2\sin^2\left(\kappa_+\hat{p}-\kappa_-\hat{x}\right), \nonumber \\
    \hat{Q}_G &= \hat{U}_G^{\dag} \hat{Q}_0 \hat{U}_G, \nonumber \\
    \hat{Q}_j &= 2 \sin^2(a \hat{x}) + 2 \sin^2(b \hat{p}),
\end{align}
which are the operators whose ground states are, respectively, GKP qubit state $|1_L\rangle$, symmetrical GKP states $|0_S\rangle$ and $|1_S\rangle$ on a square grid, hexagonal GKP on a triangular grid with $\kappa_{\pm} = \sqrt{\frac{\pi}{8}}(3^{1/4} \pm 3^{-1/4})$ \cite{grimsmo2020}, a GKP state transformed by an arbitrary \new{unitary} Gaussian operation $\hat{U}_G$ to fit into any desired encoding, and, finally, a general GKP type state on a grid with arbitrary spacing. Operators for qubit states 0 and 1 generally differ by displacement in $\hat{x}$, symmetrical grid is then obtained by squeezing \cor{\cite{supplement}}. The last operator stands aside because it cannot be always obtained from the others by a Gaussian unitary transformation and therefore does not always have a zero eigenvalue, but it is relevant for scenarios in which the grid is affected by non-unitary transformations, such as in the case of losses.
Among these new operators, $\xi_{s0}$ plays a prominent role - since the operator is symmetric, it has identical properties in both quadratures and therefore makes it easier to compare it to the existing figures of merit, which \cor{is what} we are going to do now.

\new{The GKP squeezing is directly tied to the \emph{grid squeezing} proposed by \cite{duivenvoorden2017,weigand2018}, which is also expressed as the function of stabilizers. For any quadrature operator $\hat{q}$ and the corresponding displacement operator $e^{-i u \hat{q}}$ that translates the conjugate quadrature by a grid constant $u$, the grid squeezing is defined as}
\begin{equation}\label{BTsqueezing}
    \Delta^2_{q,u} = -\frac{4}{u^2}\ln\left| \langle e^{-iu\hat{q}} \rangle \right|.
\end{equation}
\new{Please note that we have deliberetaly changed the notation from \cite{duivenvoorden2017,weigand2018} to allow for easier discussion. The GKP squeezing can be directly expressed as a function of grid squeezing values for quadratures $\hat{x}$ and $\hat{p}$:}
\begin{equation}\label{xidelta}
    \xi_{s0} = 2- e^{-\frac{\pi}{2}\Delta_{x,\sqrt{2\pi}}^2} - e^{-\frac{\pi}{2}\Delta_{p,\sqrt{2\pi}}^2}
    \approx \frac{\pi}{2} \Delta_{x,\sqrt{2\pi}}^2 + \frac{\pi}{2} \Delta_{p,\sqrt{2\pi}}^2,
\end{equation}
where the approximation holds for states with high squeezing.
\cor{This correspondence can be used to gain some insight about the suitability of the respective states for error correction \cite{bourassa2021}. The exact equivalence will need to be derived by further research, but we can expect -8.69 dB  (-5.06 dB) of GKP squeezing to be sufficient (necessary) for fault tolerance \cite{supplement}. These thresholds are invariant under Gaussian operations if they are accompanied by suitable transformation of the grid. Sufficient amount of GKP squeezing therefore indicates that the state can be transformed into an error correctable form by Gaussian transformations.}

\begin{figure}
    \includegraphics[width=0.95\linewidth]{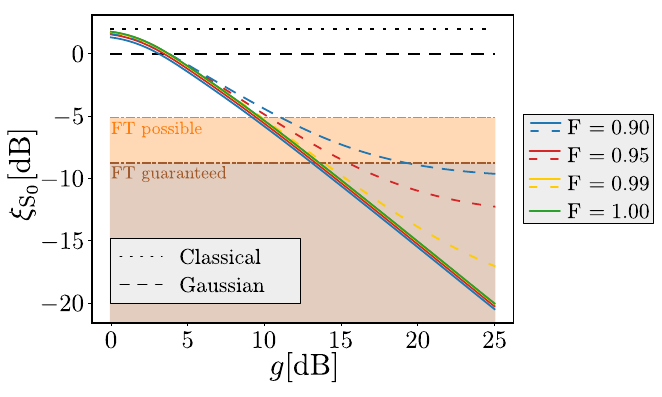}
    \caption{The range of GKP squeezing $\xi_{s0}$ for quantum states that have fidelity $F$ with an approximate state (\ref{state0g}) relative to quadrature squeezing $g$ (see the legend). Solid lines show the lower bound, dashed lines the upper bound. \cor{The black lines represent the classical and Gaussian thresholds. Background colors mark areas in which fault tolerance is guaranteed (brown) and possible (orange).}}
    \label{figFG}
\end{figure}

\new{The GKP squeezing also directly relates to the fidelity with which it is possible to prepare the approximate GKP state}
\begin{equation}\label{state0g}
    |0_{s,g}\rangle \propto \sum_{s \in Z} e^{-\frac{g}{2} (s \sqrt{2\pi})^2} \frac{1}{(\pi g)^{1/4}}\int_{-\infty}^{+\infty} e^{-\frac{(x-s\sqrt{2\pi})^2}{2 g}} |x\rangle dx,
\end{equation}
\cor{with grid squeezing $\Delta_{x,\sqrt{2\pi}}^2=\Delta_{p,\sqrt{2\pi}}^2= g$, which is equal to the normalized quadrature variance} of the individual squeezed states forming the superposition. For such state, the GKP squeezing can be found analytically as
\begin{equation}\label{}
    \xi_{s0,g,F=1} = \langle 0_{s,g}|\hat{Q}_{s0}|0_{s,g}\rangle = 2 - 2e^{-\frac{\pi g}{2}}.
\end{equation}
\new{The linearity of the operator and the bounded range of its eigenvalues guarrantee that, for any state that has fidelity $F = f$ to approximate state (\ref{state0g}) with parameter $g$, the GKP squeezing is bounded by:  }
\begin{equation}\label{}
    f \xi_{s0,g,F=1} \le \xi_{s0,g,F=f} \le f \xi_{s0,g,F=1} + 4(1-f).
\end{equation}
\cor{See \cite{supplement} for details. The upper and lower bounds are shown in Fig.~\ref{figFG}. The spread suggests that squeezing might be a more suitable figure of merit for evaluation of quantum states than the fidelity. }

The operators (\ref{Q0}) and (\ref{Qother}) are hermitian and can be, in principle, directly measured. \cor{Interestingly, for the sake of evaluation,} the measurement can be replaced by direct measurements of two quadratures. From their statistics it is now possible to obtain the values of sine and cosine functions, together with their respective error bars, and use them to construct the value of GKP squeezing. Any value smaller than one witnesses the non-Gaussian nature of the state and low enough value ensures fault tolerance. \cor{The measurement is also much more feasible than quantum tomography required for evaluating the fidelity.}

\cor{Furthermore, when trying to assess non-Gaussian nature of experimental data, one can calculate the GKP squeezing pertinent to any operator from (\ref{Q0}) and (\ref{Qother}) and look for the lowest value:}
\begin{align}\label{xigeneral}
\xi_{\mathrm{opt}} &= 2\min\bigl\langle \sin^2(c_{11}\hat{x} +  c_{12}\hat{p}+d_1) \nonumber \\
&+ \sin^2(c_{21}\hat{x}+c_{22}\hat{p}+d_2)\bigr\rangle,
\end{align}
\new{where the minimization is taken over the vector of parameters $(c_{11},c_{12},c_{21},c_{22},d_1,d_2)$. While the coefficients $c_{ij}$ can be, in principle, arbitrary, they should form a symplectic matrix multiplied by $\sqrt{\pi/2}$ for GKP states.
In all cases, the mean value (\ref{xigeneral}) can be converted to linear combination of mean values of displacement operators, which can be evaluated efficiently \cite{provaznik2022}. The minimum (\ref{xigeneral}) then represents the best achievable GKP squeezing and the optimal vector determines the relevant grid. Taking
$M_{\mathrm{GKP}} = -\ln \xi_{opt}$ now gives us an operationally defined convex monotone of non-Gaussianity specific to GKP states \cite{chitambar2019,lachman2019,filip2011}.
}

On a theoretical side, { any of operators (\ref{Q0}) and (\ref{Qother}) }can be used as a straightforward cost function for numerical optimization of state preparation protocols. The optimization numerically simulates the experimental circuits planned for the state preparation for various parameters of the setups with the goal of finding parameters for which the produced quantum state has the required properties. \cor{ Such properties can be efficiently evaluated by GKP squeezing. 
For example, it can be quickly seen that the breeding protocol \cite{weigand2018}, which is often cited as a deterministic way to prepare GKP states, actually requires the the post-selection to work \cite{supplement}.}

\begin{figure}
    \includegraphics[width=0.95\linewidth]{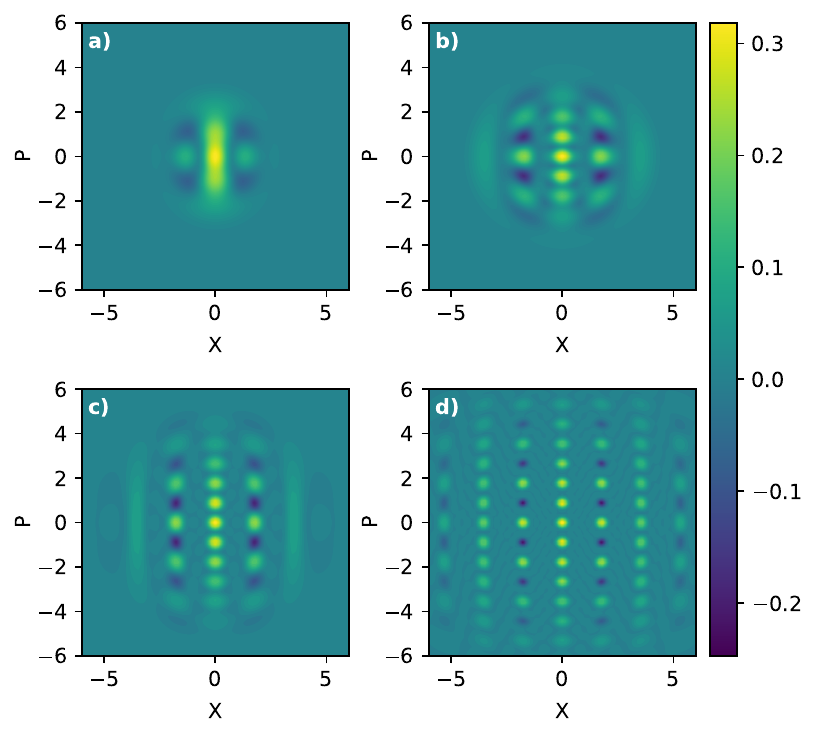}
    \caption{Wigner functions of the lowest eigenvalue eigenstates of $\hat{Q}_0$ restricted to dimension a) $\mathcal{N} = 5$, b) $\mathcal{N} =10$, c) $\mathcal{N} =20$, and d) $\mathcal{N} =50$.    }
    \label{fig1}
\end{figure}

\begin{figure}
    \includegraphics[width=0.95\linewidth]{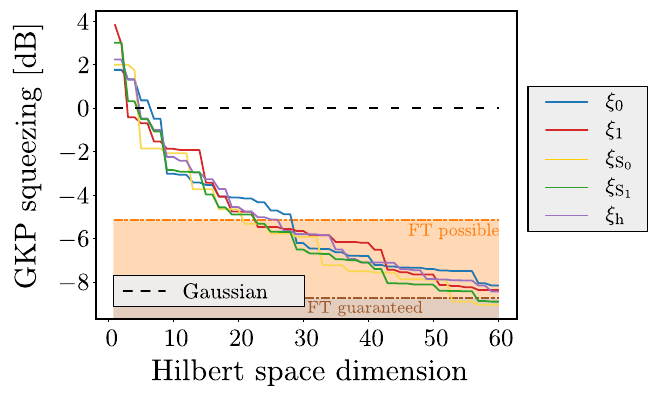}
    \caption{Maximal GKP squeezing (\ref{xi gkp}) in various topologies (see the legend) achievable for quantum states from a Hilbert space with dimension $\mathcal{N}$. \cor{ The black line represent the Gaussian threshold. Background colors mark areas in which fault tolerance is guaranteed (brown) and possible (orange).}
        }
    \label{fig2}
\end{figure}

The operators (\ref{Q0}) and (\ref{Qother}) can also give us a valuable insight on how the GKP squeezing of quantum states depends on their stellar rank \cite{chabaud2020}. We can see that by taking any one of the operators in Fock representation and turning it into a square Hermitian matrix by projecting it on a finite dimensional Hilbert space.
{The most straightforwad way of doing that relies on decomposing the operator into a sum of displacement operators (\ref{stabilizers}), \cor{constructing} their matrices on a larger Hilbert space, and then truncating them \cite{provaznik2022,supplement}}. The minimal eigenvalue of this matrix now determines the maximal achievable GKP squeezing and the corresponding eigenstate is the optimal quantum state. Fig.~\ref{fig1} shows few examples of quantum states found in this way for operator (\ref{Q0}) limited to several chosen dimensions. \new{We can also quickly compare the maximal achievable GKP squeezing for different operators (\ref{Qother}) to see which one of them is best suitable for state preparation, see  Fig.~\ref{fig2}. It can be seen that the GKP state $|1\rangle$ is the first to show a non-Gaussian behavior, already for $\mathcal{N} = 3$ corresponding to maximal photon number 2. The symmetrical GKP states $s0$ and $s1$  can be prepared slightly more easily than the other forms, but the difference is not large and the Hilbert space dimension related to the stellar rank is the determining factor.}
\begin{figure}
    \includegraphics[width=0.95\linewidth]{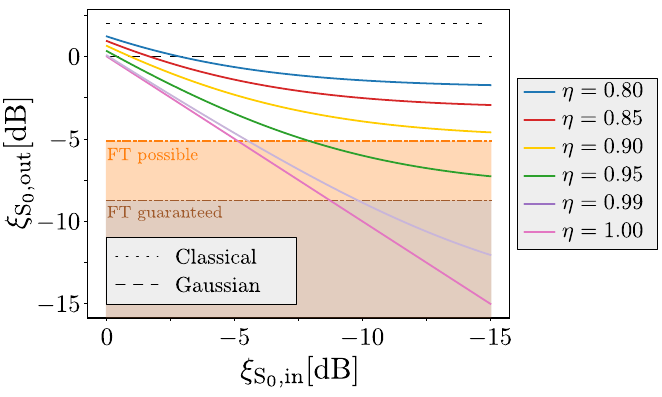}
    \caption{The GKP squeezing $\xi_{s0,\mathrm{out}}$ after decoreherence relative to the GKP squeezing before decoherence $\xi_{s0,\mathrm{in}}$ for various lossy channels (see the legend).  The black lines represent the classical and Gaussian thresholds. Background colors mark areas in which fault tolerance is guaranteed (brown) and possible (orange).}
    \label{fignoise}
\end{figure}

The GKP squeezing also offers a straightforward way to evaluate effects of the dominant forms of decoherence suffered by GKP states - Gaussian loss and additive noise. Both of these can be efficiently described with help of Heisenberg picture; the quadrature operators of the mode after the decoherence evolution are given in terms of the initial quadrature operators \new{ as $\hat{x}_{\mathrm{out}} = \sqrt{\eta}\hat{x}_{\mathrm{in}} + \hat{x}_V$ and $\hat{p}_{\mathrm{out}} = \sqrt{\eta}\hat{p}_{\mathrm{in}} + \hat{p}_V$, where $\eta $ represents the intensity loss coefficient and  $\hat{x}_V$ and $\hat{p}_V$ represent the quadrature operators of the effective environmental mode with variance $V = \langle \hat{n} \rangle + \frac{1-\eta}{2}$, where $\langle \hat{n} \rangle$ is the number of added thermal photons. The GKP squeezing after such channel can be expressed as}
\begin{align}\label{}
    \xi_{s0,\mathrm{out}} &=  2\gamma \left\langle \sin^2\Bigl(\sqrt{\frac{\pi\eta}{2}} \hat{x}_{\mathrm{in}}\Bigr)\right\rangle + 2\gamma \left\langle\sin^2\Bigl(\sqrt{\frac{\pi\eta}{2}}\hat{p}_{\mathrm{in}}\Bigr)\right\rangle \nonumber \\
    &+ 2-2\gamma,
\end{align}
where $\gamma = e^{-\pi V}$ \cite{supplement}. The two main effects of decoherence are the scaling of the grid and addition of a noise floor. The scaling of the grid can be ignored in single quadrature measurements, because it can be compensated by Gaussian phase sensitive amplification or by transforming the data. In this way, the purely lossy channel with $\eta$ can be converted to purely noisy channel with $\eta =1$ and $V= (1-\eta)/(2\eta)$. \cor{The effect is illustrated in Fig.~\ref{fignoise}, where we can see that roughly 10$\%$ losses can be expected to prevent fault tolerance.} For the symmetrical GKP squeezing, the two quadratures are affected in the similar manner. This is not true in general, different topologies behave differently under decoherence. It \cor{could be} therefore practical to take this into consideration when designing experiments \cite{filip2013,lejeannic2018}.


In summary, from their very introduction it was realized that GKP states are simultaneous eigenstates of the commuting stabilizer operators. However, these operators, individually representing displacements in two quadratures, were always treated \new{independently}. Consequently, the figures of merit, apart from the fidelity to a finite energy approximate states, were also mostly focused on \new{their independent properties}, which led to difficult trade-offs, as each stabilizer could be, on its own, saturated by a Gaussian state. We have shown that it is possible to merge the stabilizers into a single positive semi-definite Hermitian operator. In any quantum state, the mean value of this operator can be interpreted as the amount of GKP squeezing that is a monotone for the particular non-Gaussianity relevant to GKP states. At the same time, the mean value directly depends on the values of stabilizers, which allows us to keep the methods of characterization developed so far. The new operator can be feasibly evaluated for experimentally prepared states as it requires only measurement in two orthogonal bases, offers a easy-to-compute cost function for the numerical methods employed in optimization of state preparation circuits, and its eigenstates in any given dimension are the best possible approximations of GKP states. The operator can be also treated as Hamiltonian of \new{a} system for which the GKP states arise naturally as its ground states.



\begin{acknowledgements}
We acknowledge Grant No. 22-08772S of the Czech Science Foundation, the European Union's HORIZON Research and Innovation Actions under Grant Agreement no. 101080173 (CLUSTEC), and funding from Ministry of Education, Youth and Sport of the Czech Republic (CZ$.02.01.01/00/22\_008/0004649$).
We would also like to thank \v{S}imon Br\"auer  and Jan Provazn\'{i}k for the help with visualization of the results, and the anonymous referees for great discussion and insightful comments.
\end{acknowledgements}

\newpage
\onecolumngrid
\section{Appendix: Classical and Gaussian thresholds}
Let us consider a general operator of form
\begin{equation}\label{qg}
    \hat{Q}_j = 2 \sin^2(a \hat{x}) + 2 \sin^2(b \hat{p}),
\end{equation}
where $\hat{x}$ and $\hat{p}$ are quadrature operators with $[\hat{x},\hat{p}] = i$, and $a,b$ are real positive numbers, and derive the minimal mean values that can be obtained by Gaussian and classical states. Let us start with the bound for classical states, which are defined as states that can be represented as a classical mixture of coherent states. Since the mean value is a linear functional of a density matrix, the minimal value for classical states will be the mean value of the operator in a specific coherent state and since the operator (\ref{qg}) is symmetric with respect to both $\hat{x}$ and $\hat{p}$, the minimizing state will have these symmetries. The minimum over the classical states will then be found for the vacuum state $|\mathrm{vac}\rangle$ and it will be equal to
\begin{align}\label{}
    M_C = \langle \mathrm{vac}|\hat{Q}_j |\mathrm{vac}\rangle
         = \frac{1}{\sqrt{\pi}}\int_{-\infty}^{+\infty} [2 \sin^2(a q) + 2 \sin^2(b q)]e^{-q^2}dq = 2 - e^{-a^2} - e^{-b^2}.
\end{align}
Let us now turn our focus to the Gaussian states, defined as mixtures of Gaussian states. Again, due to the linearity of operator (\ref{qg})  it is sufficient to find a pure Gaussian state that minimizes the mean value. And due to the  symmetry considerations, such state will be a vacuum state squeezed in one of the quadratures:
\begin{equation}\label{}
    |\mathrm{Sq};g,0\rangle = (\pi g)^{-\frac{1}{4}}\int_{-\infty}^{+\infty} e^{-\frac{x^2}{2g}}|x\rangle dx.
\end{equation}
The minimum is then
\begin{align}\label{minimum}
M_G & = \min_g \langle\mathrm{Sq};g,0| \hat{Q}_j |\mathrm{Sq};g,0\rangle \nonumber \\
&= \min_g \int_{-\infty}^{+\infty} \left[ 2 \sin^2(a q)\frac{e^{-q^2/g}}{\sqrt{\pi g}} + 2 \sin^2(b q) \frac{\sqrt{g}e^{-g q^2}}{\sqrt{\pi}} \right]dq = \nonumber \\
&= \min_g \left[2 - e^{-\frac{a^2}{g}} - e^{-b^2 g} \right].
\end{align}
We can now see that, as long as $ab>\ln 2 \approx 0.69$, the minimum can be only be achieved in a limit, either $g\rightarrow 0$ or $g\rightarrow +\infty$, and that it is equal to $M_G = 1$. Note that we can also find a relaxed bound achievable by Gaussian states with finite squeezing by limiting the range over which we minimize $g$ in (\ref{minimum}). This bound could allow us to verify practical non-Gaussianity even in experimental data that do not overcome the full Gaussian bound.

\newpage
\section{Appendix: Gaussian transformations}
The different operators for GKP squeezing,
\begin{align}\label{a:Qother}
    \hat{Q}_{0} &= 2\sin^2\left(\frac{\hat{x}\sqrt{\pi}}{2}\right) + 2\sin^2(\hat{p}\sqrt{\pi}) \nonumber \\
    \hat{Q}_{1} &= 2\cos^2\left(\frac{\hat{x}\sqrt{\pi}}{2}\right) + 2\sin^2\left(\hat{p}\sqrt{\pi}\right),\nonumber \\
    \hat{Q}_{s0}& =  2\sin^2\left(\hat{x}\sqrt{\frac{\pi}{2}}\right) + 2\sin^2\left(\hat{p}\sqrt{\frac{\pi}{2}}\right), \nonumber \\
    \hat{Q}_{s1} &=  2\cos^2\left(\hat{x}\sqrt{\frac{\pi}{2}}\right) + 2\sin^2\left(\hat{p}\sqrt{\frac{\pi}{2}}\right), \nonumber \\
    \hat{Q}_h &= 2\sin^2\left(\kappa_+\hat{x}-\kappa_-\hat{p}\right) + 2\sin^2\left(\kappa_+\hat{p}-\kappa_-\hat{x}\right), \nonumber \\
    \hat{Q}_G &= \hat{U}_G^{\dag} \hat{Q}_0 \hat{U}_G, 
\end{align}
with $\kappa_{\pm} = \sqrt{\frac{\pi}{8}}(3^{1/4} \pm 3^{-1/4})$, can be all transformed into each other by Gaussian operations. Any of these operators can be expressed as a function of quadrature operators which can be arranged into a vector $\hat{\zeta} = (\hat{x},\hat{p})^T$, for example, $\hat{Q}_0 = Q_0(\hat{\zeta})$. Gaussian operations then transform the operators by transforming the vectors of quadratures:
\begin{equation}\label{}
    \hat{Q}_G = Q_G(\hat{\zeta}) = Q_0(\hat{U}_G^{\dag}\hat{\zeta}\hat{U}_G).
\end{equation}
Any Gaussian transformation can be now represented by a symplectic matrix $A$ and a vector $\alpha$ that, together, transform the vector of quadratures as
\begin{equation}\label{}
    \hat{U}_G^{\dag}\hat{\zeta}\hat{U}_G = A \hat{\zeta} + \alpha.
\end{equation}
Physically, matrix $A$ represents phase shift and squeezing, while the vector $\alpha$ describes quadrature displacement. It is now straightforward to check that the various GKP operators (\ref{a:Qother}) can be obtained from $\hat{Q}_0$ by applying Gaussian operations with
\begin{align}
A_1 = \left(\begin{array}{cc}
              1 & 0 \\
              0 & 1
            \end{array}\right), \quad\alpha_1 = \left(\begin{array}{c}
                                              \sqrt{\pi} \\
                                              0
                                            \end{array}\right) \nonumber \\
A_{s0} = \left(\begin{array}{cc}
              \sqrt{2} & 0 \\
              0 & \frac{1}{\sqrt{2}}
            \end{array}\right), \quad\alpha_{s0} = \left(\begin{array}{c}
                                              0 \\
                                              0
                                            \end{array}\right) \nonumber \\
A_{s1} = \left(\begin{array}{cc}
              \sqrt{2} & 0 \\
              0 & \frac{1}{\sqrt{2}}
            \end{array}\right), \quad\alpha_{s1} = \left(\begin{array}{c}
                                              \sqrt{\frac{\pi}{2}} \\
                                              0
                                            \end{array}\right) \nonumber \\
A_{h} =
\left(\begin{array}{cc}
              \sqrt{2}\cosh r & \frac{-\sinh r}{\sqrt{2}} \\
              -\sqrt{2} \sinh r & \frac{\cosh r}{\sqrt{2}}
            \end{array}\right),
                        \quad\alpha_h = \left(\begin{array}{c}
                                              0 \\
                                              0
                                            \end{array}\right),
\end{align}
where $r = \frac{\ln 3}{4}$. The first three operations are a simple stabilizer displacements and/or quadrature squeezing. The last operation, leading to the triangular grid, consists of quadrature squeezing a specific direction. The squeezing can be decomposed into two squeezing operations - the first one transforms the grid into a symmetrical one, as in the case of $\hat{Q}_{s0}$, while the second squeezing, this time rotated by $\frac{\pi}{4}$, changes the squares of the grid into rhombi composed of two equilateral triangles.

\newpage
\section{Appendix: Relation to existing figures of merit}
The GKP squeezing can be directly related to the two main figures of merit currently used for evaluation of GKP states. Both of these figures of merit are based on comparing the evaluated state to the approximation of the ideal GKP states found as superpositions of displaced finitely squeezed vacuum states
\begin{align}\label{approxstate}
|0_{g}\rangle \propto \sum_{s \in Z} e^{-\frac{g}{2} (2 s a)^2} |\mathrm{Sq};g,2sa\rangle = \sum_{s \in Z} e^{-\frac{g}{2} (2 s a)^2} \frac{1}{(\pi g)^{1/4}}\int_{-\infty}^{+\infty} e^{-\frac{(x-2sa)^2}{2 g}} |x\rangle dx,\nonumber \\
|1_{g}\rangle \propto \sum_{s \in Z} e^{-\frac{g}{2} (2 s a + a)^2}|\mathrm{Sq};g,2sa+a\rangle = \sum_{s \in Z} e^{-\frac{g}{2} (2 s a + a)^2} \frac{1}{(\pi g)^{1/4}}\int_{-\infty}^{+\infty} e^{-\frac{(x-2sa-a)^2}{2 g}} |x\rangle dx,
\end{align}
where $a$ determines the grid spacing and $g$ is the normalized variance of the constituent squeezed states. For the sake of comparison, we shall be mainly interested in the GKP states on the \cor{logical grid with $a = \sqrt{\pi}$ and the} symmetrical grid, for which \cor{ $a = \sqrt{\frac{\pi}{2}}$}. For evaluating the properties of these states it is important to be able to calculate the mean values of arbitrary displacement operators, $\langle e^{ib\hat{x}}\rangle$ and $\langle e^{ib\hat{p}}\rangle$. These can be evaluated with help of
\begin{align}\label{evaluating}
\langle \mathrm{Sq};g,2s_1a| e^{ib\hat{x}}|\mathrm{Sq};g,2s_2a\rangle & = \exp \left[ -\frac{b^2g}{4} - \frac{a^2 (s_1-s_2)^2}{g} + i ab(s_1+s_2) \right] \nonumber \\
\langle \mathrm{Sq};g,2s_1a+a| e^{ib\hat{x}}|\mathrm{Sq};g,2s_2a+a\rangle & = \exp \left[ -\frac{b^2g}{4} - \frac{a^2 (s_1-s_2)^2}{g} + i ab(s_1+s_2+1) \right] \nonumber \\
\langle \mathrm{Sq};g,2s_1a| e^{ib\hat{p}}|\mathrm{Sq};g,2s_2a\rangle & = \langle \mathrm{Sq};g,2s_1a+a| e^{ib\hat{p}}|\mathrm{Sq};g,2s_2a+a\rangle = \exp\left[ -\frac{(b+2as_1 - 2as_2)^2}{4g}\right].
\end{align}

\begin{figure}
    \includegraphics[width=0.9\linewidth]{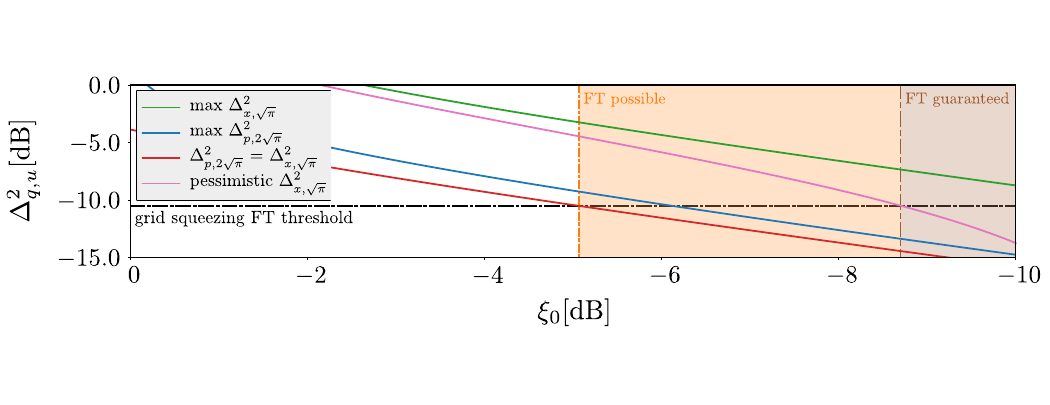}
    \caption{Achievable grid squeezing values $\Delta_{q,u}^2$ for an arbitrary quadrature $q$, relative to the GKP squeezing $\xi_{0}$ for the logical grid. Blue solid line: maximal possible value of $\Delta_{p,2\sqrt{\pi}}^2$, green solid line: maximal possible value of $\Delta_{x,\sqrt{\pi}}^2$, red solid line: symmetric scenario with $\Delta_{p,2\sqrt{\pi}}^2 = \Delta_{x,\sqrt{\pi}}^2$, magenta solid line: the pessimistic scenario for which $\Delta_{p,2\sqrt{\pi}}^2 \approx 0.045$.  The black dot-dashed line marks the -10.5 fault tolerance threshold from \cite{bourassa2021}. The vertical lines mark the approximate necessary (orange) and sufficient (brown) values for fault tolerance. }
    \label{figS1}
\end{figure}

The first figure of merit is the grid squeezing, proposed by Weigand and Terhal \cite{weigand2018}, which can be, for an arbitrary quadrature operator $\hat{q}$, obtained from the mean value of a chosen displacement operator:
\begin{equation}\label{}
   \Delta^2_{q,u} = -\frac{4}{u^2}\ln\left| \langle e^{-iu\hat{q}} \rangle \right|,
\end{equation}
where the real parameter $u$ defines the grid spacing, \cor{and relates to (\ref{approxstate}) as $u = 2 a$}. The chosen case of symmetrical grid can be obtained for $u =  \sqrt{2\pi}$. For states (\ref{approxstate}) and matching grid parameters, the grid squeezing evaluates to $\Delta_{q,u}^2 = g$ for both quadratures $\hat{x}$ and $\hat{p}$. The grid squeezing can be directly tied to the GKP squeezing under a single fair assumption. The mean value of the displacement operator is, in general, a complex number. However, the imaginary component can be removed by an additional fixed displacement, which is generally assumed in simulations and evaluations. We can therefore consider the mean value of $\langle e^{-iu\hat{q}} \rangle$ to be real and positive, which means we can use
\begin{align}
\Delta^2_{q,u} = -\frac{4}{u^2}\ln\left|\frac{\langle e^{-iu\hat{q}} + e^{iu\hat{q}}\rangle}{2} \right| =  -\frac{4}{u^2}\ln\left|\Bigl\langle 1- 2 \sin^2\Bigl(\frac{u \hat{q}}{2}\Bigr) \Bigr\rangle \right|,
\end{align}
and directly find
\begin{equation}\label{}
    \left\langle \sin^2\Bigl(\frac{u\hat{q}}{2}\Bigr) \right\rangle = \frac{1 - e^{-\frac{u^2 \Delta_{q,u}^2}{4}}}{2} \approx \frac{u^2 \Delta_{q,u}^2}{8}.
\end{equation}
We can now see that, for any grid spacing, the GKP squeezing is a \cor{sum of monotonic functions} of the two grid squeezing values and that the functions become linear in the limit of high enough squeezing. Specifically, for the case of symmetrical \cor{and logical} grids we have
\begin{align}\label{a:xidelta}
   \xi_0 &= \langle \hat{Q}_0\rangle = 2-e^{-\frac{\pi}{4}\Delta_{x,\sqrt{\pi}}^2} - e^{-\pi \Delta_{p,2\sqrt{\pi}}^2}
   \approx \frac{\pi}{4}\Delta_{x,\sqrt{\pi}}^2 +  \pi \Delta_{p,2\sqrt{\pi}}^2 \nonumber \\
   \xi_{s0} &=  \langle \hat{Q}_{s0} \rangle = 2- e^{-\frac{\pi}{2}\Delta_{x,\sqrt{2\pi}}^2} - e^{-\frac{\pi}{2}\Delta_{p,\sqrt{2\pi}}^2}
   \approx \frac{\pi}{2} \Delta_{x,\sqrt{2\pi}}^2 + \frac{\pi}{2} \Delta_{p,\sqrt{2\pi}}^2.
\end{align}
\cor{Note that using Gaussian operations to convert GKP states from one grid to another preserves the amount of GKP squeezing in the state. This is because the Gaussian operation applies a linear transformation to the quadrature operators that is inverse to the transformation of the grid parameters. Thanks to this property, fault tolerance thresholds derived for a single grid can be extended for any state with GKP squeezing. Any state that has sufficient GKP squeezing can be then converted to the form that allows error correction by Gaussian operations. Finding such thresholds is a question for future research, but we can immediately take advantage of fault tolerance thresholds derived for the states on the logical grid. In \cite{bourassa2021} it was shown that approximate states (\ref{approxstate}) are fault tolerant for $a = \sqrt{\pi}$ and $\Delta_{x,\sqrt{\pi}}^2 = \Delta_{p,2\sqrt{\pi}}^2 =  0.089$, which corresponds to -10.5 dB of grid squeezing. From (\ref{a:xidelta}) we can, for any GKP squeezing values, bound the grid squeezing values from above by}
\begin{align}
\Delta_{x,\sqrt{\pi}}^2 &\le -\frac{4}{\pi} \ln(1-\xi_0), \\ \nonumber
\Delta_{p,2\sqrt{\pi}}^2 &\le -\frac{1}{\pi} \ln(1-\xi_0), \\ \nonumber
\Delta_{x,\sqrt{2\pi}}^2 &\le -\frac{2}{\pi} \ln(1- \xi_{s0}), \\ \nonumber
\Delta_{p,\sqrt{2\pi}}^2 &\le -\frac{2}{\pi} \ln(1- \xi_{s0}).
\end{align}
\cor{This bound is obtained by assuming the second grid squeezing value equal to zero. We can now see that GKP squeezing of $\xi_0 = 0.068$, corresponding to -11.69 dB, ensures that the grid squeezing values are both at least 10.5 dB, which should ensure fault tolerance. However, this bound can be made less strict by realizing that logical GKP states are preserved under the action of squeezing and phase shift. GKP state $|\tilde{0}_L\rangle$ with some $\Delta_{x,\sqrt{\pi}}^2 = g_1$ and $\Delta_{p,2\sqrt{\pi}}^2 = g_2$ can be, by Gaussian operations, converted to another state  $|\tilde{0}_L\rangle$ with $\Delta_{x,\sqrt{\pi}}^2 = 2g_2$ and $\Delta_{p,2\sqrt{\pi}}^2 =  \frac{g_1}{2}$. We can therefore consider a pessimistic scenario, in which $\Delta_{p,2\sqrt{\pi}}^2 \approx 0.045$, which is only slightly worse than -13.5 dB. This surpasses the fault tolerant threshold of -10.5 dB, but such state can not be converted to a state with sufficiently low $\Delta_{x,\sqrt{\pi}}^2$. In this scenario, to have both grid squeezing variances surpassing the fault tolerance threshold, we need GKP squeezing of $\xi_0 = 0.135$, or -8.69 dB. We can also consider the optimistic scenario in which $\Delta_{x,\sqrt{\pi}}^2 = \Delta_{p,2\sqrt{\pi}}^2 = 0.089$ and which corresponds to GKP squeezing of $\xi_0 = 0.312$, or -5.06 dB. This value does not guarantee fault tolerance, but it marks the area in which it it s possible. This relation is explicitly shown in Fig.~\ref{figS1}.  Note that the equivalence is not absolute, since the threshold was derived specifically for states (\ref{approxstate}) rather than for states defined by their grid squeezing, but it can be considered a reasonable starting estimate.}

\begin{figure}
    \includegraphics[width=0.9\linewidth]{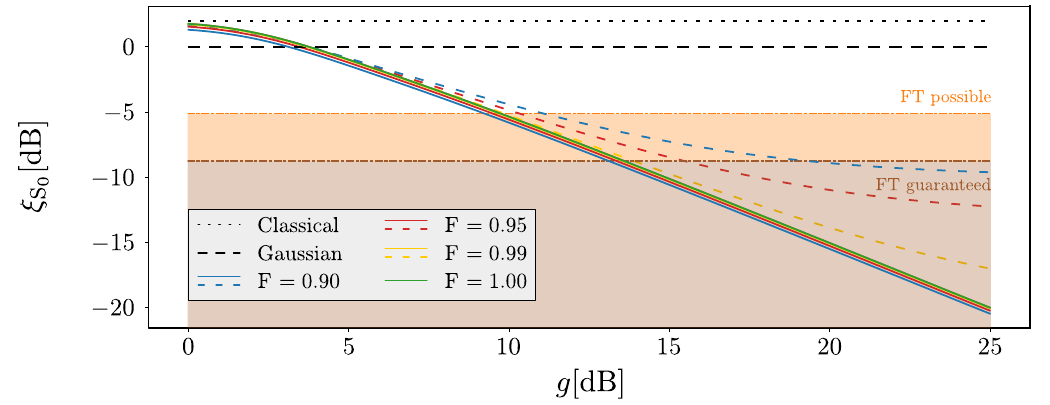}
    \caption{The colored lines depict the upper (dashed lines) and lower (solid lines) bounds on GKP squeezing $\xi_{s0}$ for quantum states with specific values of fidelity with the target state $|0_g\rangle$ for \cor{$a = \sqrt{\frac{\pi}{2}}$}. Please see the legend for differentiation. The black lines represent the classical and Gaussian thresholds. The color filled areas mark the approximate necessary (orange) and sufficient (brown) thresholds for fault tolerance. }
    \label{figS2}
\end{figure}

The second figure of merit commonly used in evaluation of GKP states is the fidelity with the ideal approximated state (\ref{approxstate}). This evaluation always yields two relevant values - the squeezing of the Gaussian states in the superposition given by the parameter $g$, and the fidelity. The interplay between these two parameters is not trivial. For evaluation of GKP squeezing, we can take an arbitrary GKP squeezing operator $\hat{Q}_{j}$, (\ref{qg}), and find its value for any specific state (\ref{approxstate}) in the form
\begin{equation}\label{}
    \xi_{j,g,F=1} = \langle 0_g|\hat{Q}_j|0_g\rangle,
\end{equation}
which can be performed analytically. From the linearity of the operator and from the fact that its eigenvalues are from the interval $[0,4]$ then immediately follows that any approximation of the state with a different fidelity $F = f$ must adhere to
\begin{equation}\label{}
    f \xi_{j,g,F=1} \le \xi_{j,g,F=f} \le f \xi_{j,g,F=1} + 4(1-f).
\end{equation}
This provides us with the bound on the GKP squeezing both from above and from below for the particular values of fidelity $f$ and squeezing $g$. It also provides bounds on the fidelity $f$ and squeezing $g$ for any GKP squeezing of the state. The relation is illustrated in Fig.~\ref{figS2}. Interestingly, we can see that the targeted squeezing $g$ is actually much more important than the value of fidelity with the state. \cor{This suggests that, for state preparation, squeezing might be a better figure of merit than the fidelity.}

\newpage
\section{Appendix: Evaluation of state preparation schemes}
The schemes proposed for preparation of GKP states are varied, but they can be broadly separated into three approaches. The first one attempts to prepare a specific target superposition of photon number states
\begin{equation}\label{focksuperposition}
    |\psi_{T}\rangle = \sum_{k} c_k |\hat{n}=k\rangle.
\end{equation}
This is is the most common approach in quantum optical schemes that utilize either boson sampling or breeding of approximate superposed coherent states. The quality of these states can be evaluated by the values of the stabilizers or by the fidelity with some approximate GKP state. However, these figures of merit always involve some trade-offs, either between the values of two different stabilizers, or between the fidelity and the peak squeezing of the approximate GKP states. This complicates matters when it comes to actually finding the ideal coefficients in the superposition (\ref{focksuperposition}). On the other hand, for any kind of GKP squeezing operator (\ref{qg}), the optimal finite superposition of photon number states that minimizes its value can be always found as a minimum value eigenstate of a truncated operator
\begin{equation}\label{Qgnmax}
    \hat{Q}_{j,N_{\mathrm{max}}} = \left[\sum_{k=0}^{N_{\mathrm{max}}}|\hat{n}=k\rangle\langle\hat{n}=k|\right] \hat{Q}_{j}
    \left[\sum_{k=0}^{N_{\mathrm{max}}}|\hat{n}=k\rangle\langle\hat{n}=k|\right],
\end{equation}
where $N_{\mathrm{max}}$ is the maximal considered photon number in the superposition that relates to the state's stellar rank. The respective minimum eigenvalue is then equal to the optimal value of the GKP squeezing. This allows us both to effectively find the required states and to evaluate different kinds of GKP operators to find the one that leads to the best squeezing under the chosen conditions.

Naturally, the most important part of this evaluation is having the correct form of operator (\ref{Qgnmax}). This is not trivial, since operators $\hat{x}$ and therefore $\sin(\hat{x})$ span the whole infinite dimensional Hilbert space of the harmonic oscillator. However, we can take advantage of
\begin{equation}\label{sinexpansion}
    \sin^2(a\hat{x}) = \frac{1}{4}\left( 2 - e^{-i 2 a \hat{x}} - e^{i 2 a \hat{x}}\right)
\end{equation}
and expand the operator (\ref{qg}) as a linear combination of displacement operators. Their explicit form can now be found by several methods, ranging from analytical formulas to effective numerical methods \cite{provaznik2022}. The most straightforward approach is to use standard numerical tools to generate the matrix of the displacement operator on a larger dimension, such as $N= 10 N_{\mathrm{max}}$, and then cut it to size. The actual required size of the larger Hilbert space depends on the value of $a$, but this approach generally can produce the exact form of the operator with high precision and at reasonable speeds \cite{provaznik2022}. The form of (\ref{sinexpansion}) then ensures Hermiticity of operator (\ref{Qgnmax}) and its positive eigenvalues.

Note that this approach can be applied to any function $\sin^2(a \hat{x} + b\hat{p})$, \cor{arising}, for example, by applying a general Gaussian unitary operation to the GKP squeezing operator. The argument of such general function can be also written as
\begin{equation}\label{}
   a \hat{x} + b \hat{p} = z \hat{x} \cos\phi  + z \hat{p}\sin\phi  = z \hat{x}(\phi),
\end{equation}
where $z = \sqrt{a^2+b^2}$ and $\phi = \mathrm{atan}\frac{b}{a}$. The operator can then be again expressed as linear combination of single quadrature displacement operators (\ref{sinexpansion}). This allows straightforward evaluation of GKP squeezing on any kind of grid, even though for highly warped cases, in which $z$ is large, the requirement on Hilbert space dimension may be considerable. However, this would also be the case for the alternative evaluation methods.

\begin{figure}
    \includegraphics[width=0.9\linewidth]{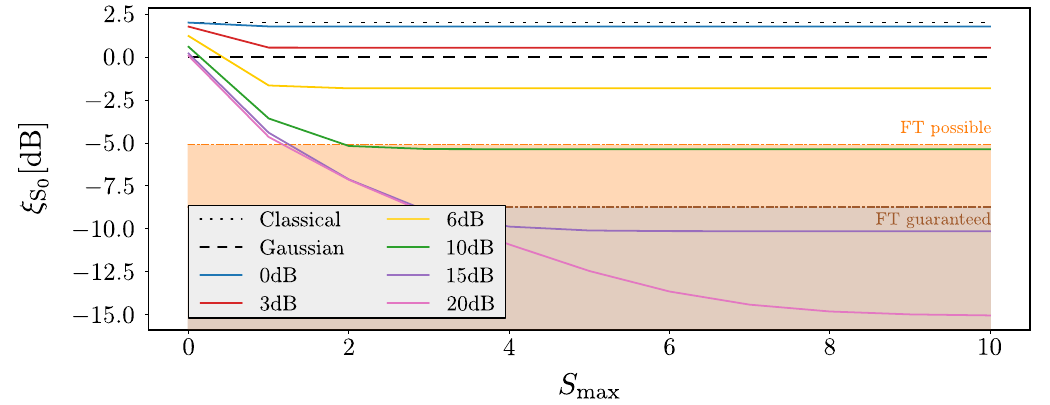}
    \caption{GKP squeezing of approximate states (\ref{finiteS}) relative to the initial squeezing $g$ and the number of peaks related to $S_{\mathrm{max}}$. The colored lines depict the GKP squeezing, please see the legend for differentiation. The black lines represent the classical and Gaussian thresholds. The color filled areas mark the approximate necessary (orange) and sufficient (brown) thresholds for fault tolerance. }
    \label{figS3}
\end{figure}

The second way to prepare the GKP states is by starting from a squeezed state and then applying superpositions of displacement operators, directly creating a finite superposition of squeezed states looking as
\begin{equation}\label{finiteS}
    |0_{g}\rangle \propto \sum_{s = - S_{\mathrm{max}}}^{S_{\mathrm{max}}} e^{-\frac{g}{2} (2 s a)^2} |\mathrm{Sq};g,2sa\rangle = \sum_{s = - S_{\mathrm{max}}}^{S_{\mathrm{max}}} e^{-\frac{g}{2} (2 s a)^2} \frac{1}{(\pi g)^{1/4}}\int_{-\infty}^{+\infty} e^{-\frac{(x-2sa)^2}{2 g}} |x\rangle dx.
\end{equation}
The state then approaches the ideal GKP state as the number of peaks, given as $2 S_{\mathrm{max}}+1$, increases.
This approach is best suited for systems in which the state of the harmonic oscillator can be coupled to a two level system, such as trapped ion or a superconducting qubit. The quality of the finite superposition (\ref{finiteS}) can be quickly evaluated with help of (\ref{evaluating}) and the dependence is, for various values of initial squeezing $g$ shown in Fig.~\ref{figS3}.

Finally, let us briefly discuss the breeding protocol put forth in \cite{weigand2018}, which is often mistakenly presented as a deterministic way to prepare GKP states from an ensemble of squeezed superposed coherent states. A single step of the protocol consists of mixing two approximate GKP states on a balanced beam splitter, following by measuring one of the output modes and performing a suitable corrective operation.
If we assume the deterministic nature, we should be able to evaluate the protocol in the Heisenberg picture. Using this approach, we can express the GKP squeezing of a single mode on an arbitrary grid after a single step as
\begin{align}\label{}
\xi_{j,\mathrm{out}} &= 2\left\langle \sin^2(a\hat{x}_{\mathrm{out}}) + \sin^2(b\hat{p}_{\mathrm{out}})\right\rangle \nonumber \\
&= 2\left\langle \sin^2\Bigl(a\frac{\hat{x}_{1}+\hat{x}_2}{\sqrt{2}}\Bigr) + \sin^2\Bigl(b\frac{\hat{p}_{1}+\hat{p}_2}{\sqrt{2}}\Bigr)\right\rangle \nonumber
\end{align}
where $\hat{x}_1, \hat{p}_1, \hat{x}_2,\hat{p}_2$ are the quadrature operators of the input modes that are assumed to be both in identical approximative GKP states. Measuring the second output mode yields value $p_M = \frac{\hat{p}_2 - \hat{p}_1}{\sqrt{2}}$, which allows us to apply suitable feed-forward to the $\hat{p}$ quadrature and rewrite the GKP squeezing as
\begin{align}
\xi_{j,\mathrm{out}} &= 2\left\langle
\sin^2\Bigl(a\frac{\hat{x}_{1}}{\sqrt{2}}\Bigr)\cos^2\Bigl(a\frac{\hat{x}_2}{\sqrt{2}}\Bigr) +
\cos^2\Bigl(a\frac{\hat{x}_{1}}{\sqrt{2}}\Bigr)\sin^2\Bigl(a\frac{\hat{x}_2}{\sqrt{2}}\Bigr) +
\frac{1}{2}\sin(a\sqrt{2}\hat{x}_1)\sin(a\sqrt{2}\hat{x}_2) + \sin^2(b\sqrt{2}\hat{p}_1)
\right\rangle \nonumber \\
& = 4\left\langle \sin^2\Bigl(a\frac{\hat{x}_{1}}{\sqrt{2}}\Bigr)\right\rangle
\left\langle 1- \sin^2\Bigl(a\frac{\hat{x}_{1}}{\sqrt{2}}\Bigr)\right\rangle
+ 2\left\langle\sin^2(b\sqrt{2}\hat{p}_1)\right\rangle,
\end{align}
where, in the last line, we have assumed that the two input modes are independent and in the same state with an even wave function. We can now see that the single breeding step can not improve the GKP squeezing, unless it was initially rather low. The main effect is actually just the scaling of the grid, also achievable by a squeezing operation. However, this does not mean that the protocol does not work. Only that the initial assumption on its deterministic nature was wrong and that the post-selection step is a critical component of the protocol and cannot be dismissed, even though its success rate may approach one.

\newpage
\section{Appendix: Loss and Gaussian noise}
Let us again consider the general operator $\hat{Q}_j$, (\ref{qg}), and evaluate how it is transformed under the action of common sources of decoherence - loss and additive noise. Since these two processes behave in a similar fashion, we can consider a general case in which loss and noise are applied simultaneously. Such channel can be described in the Heisenberg picture by transformation relations for quadrature operators. The output quadrature operators depend on the input quadratures as
\begin{equation}\label{}
    \hat{x}_{\mathrm{out}} = \sqrt{\eta}\hat{x}_{\mathrm{in}} + \hat{x}_{V_x}, \quad \hat{p}_{\mathrm{out}} = \sqrt{\eta}\hat{p}_{\mathrm{in}} + \hat{p}_{V_p},
\end{equation}
where $\eta$ is the intensity transmission coefficient of the channel, and $\hat{x}_{V_x}$ and $\hat{p}_{V_p}$ are the effective quadrature operators for the environmental mode.  Their subscripts denote the amounts of Gaussian noise that is added to the respective quadrature. For a symmetric channel with both loss and noise, $V_x = V_p = \langle \hat{n} \rangle + \frac{1-\eta}{2}$, where $\langle \hat{n}\rangle $ is the average number of excitations added to the mode. Only loss or only noise can be then obtained by setting $\langle \hat{n}\rangle = 0$ and $\eta = 1$, respectively.

\begin{figure}
    \includegraphics[width=0.9\linewidth]{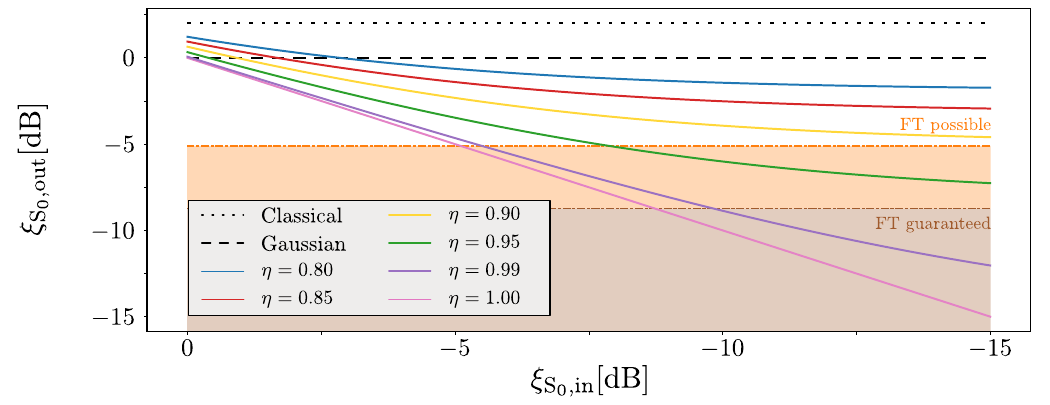}
    \caption{The colored lines show the effective GKP squeezing $\xi_{S0,\mathrm{out}}$ for a state with initial GKP squeezing $\xi_{S0,\mathrm{in}}$ affected by a lossy channel with transmission coefficient $\eta$ measured in the scaled basis. Please see the legend for differentiation. The black lines represent the classical and Gaussian thresholds. The color filled areas mark the approximate necessary (orange) and sufficient (brown) thresholds for fault tolerance.}
    \label{figS4}
\end{figure}

After such channel, the GKP squeezing can can be expressed as
\begin{align}\label{decoherence}
    \xi_{j,\mathrm{out}} = \langle \hat{Q}_j\rangle_{\mathrm{out}} &=
    \left\langle 2 \sin^2[a (\sqrt{\eta}\hat{x}_{\mathrm{in}} +\hat{x}_{V_x})] + 2 \sin^2[b (\sqrt{\eta}\hat{p}_{\mathrm{in}} + \hat{p}_{V_2})]  \right\rangle \nonumber \\
    & = \left\langle 1- \cos[2a (\sqrt{\eta}\hat{x}_{\mathrm{in}} +\hat{x}_{V_x})] + 1-\cos[2b (\sqrt{\eta}\hat{p}_{\mathrm{in}} + \hat{p}_{V_p})] \right\rangle \nonumber \\
    & = 2 - \langle \cos(2a\sqrt{\eta} \hat{x}_{\mathrm{in}})\rangle \langle\cos(2a\hat{x}_{V_x})\rangle +
    \langle \sin(2a \sqrt{\eta}\hat{x}_{\mathrm{in}})\rangle \langle\sin(2a\hat{x}_{V_x})\rangle \nonumber\\
    &\quad - \langle\cos(2b \sqrt{\eta}\hat{p}_{\mathrm{in}})\rangle\langle\cos( 2b\hat{p}_{V_p})\rangle + \langle\sin(2b \sqrt{\eta}\hat{p}_{\mathrm{in}})\rangle\langle\sin( 2b\hat{p}_{V_p})\rangle.
\end{align}
In the cases of loss, additive Gaussian noise, and their combinations, the state  of the environment is symmetrical with respect to both quadratures. The terms containing the sine functions therefore reduce to zero. On the other hand, the terms containing the cosine functions of the environmental modes can be directly evaluated to be
\begin{align}\label{}
    \gamma_x = \langle\cos(2a\hat{x}_{V_x})\rangle =  \frac{1}{\sqrt{\pi 2 V_x}} \int_{-\infty}^{+\infty}\cos(2a q)e^{-\frac{q^2}{2 V_x}}dq = e^{-2 a^2 V_x}, \nonumber \\
    \gamma_p = \langle\cos( 2b\hat{p}_{V_p})]\rangle = \frac{1}{\sqrt{\pi 2 V_p}} \int_{-\infty}^{+\infty}\cos(2b q)e^{-\frac{q^2}{2 V_p}}dq = e^{-2 b^2 V_p}.
\end{align}
The mean value (\ref{decoherence}) can be now expressed as
\begin{align}\label{}
    \langle \hat{Q}_j\rangle_{\mathrm{out}} & = 2 - \gamma_x \langle \cos(2a\sqrt{\eta} \hat{x}_{\mathrm{in}})\rangle
    - \gamma_p \langle\cos(2b \sqrt{\eta}\hat{p}_{\mathrm{in}})\rangle \nonumber \\
    & = 2\gamma_x \langle \sin^2(a\sqrt{\eta} \hat{x}_{\mathrm{in}})\rangle + 2\gamma_p \langle\sin^2(b \sqrt{\eta}\hat{p}_{\mathrm{in}})\rangle + 2-\gamma_x-\gamma_p.
\end{align}
It is apparent that there are two main consequences of the decoherence. The losses affect the grid spacing; the states shrink towards the point of origin of the phase space which means that their GKP squeezing corresponds to GKP squeezing of the original states on an inflated grid. This naturally limits the achievable GKP squeezing, because even though the original states can be prepared on the inflated grid, their GKP squeezing can not be arbitrarily low, because the operators no longer commute. Both losses and added noise then add a flat additive noise that limits the GKP squeezing even further. Note that losses can be converted to the additive noise, either by performing active amplification, or, in the case only a single quadrature measurement is relevant, by scaling the measured quadrature. In a similar manner, effects of additive noise can be represented by loss. Let us therefore consider only loss, but express it as additive noise. In this frame, a purely lossy channel can be represented by a purely noisy channel with $V_x = V_p = (1-\eta)/(2\eta)$. Dependance of the GKP squeezing for such channel is depicted in Fig.~\ref{figS4}. \cor{We can see that 10$\%$ losses, which roughly correspond to adding 0.056 thermal photons, can be expected to prevent fault tolerance for any quantum state}.


\end{document}